\documentclass[11pt,twoside]{article}
\usepackage{asp2004}
\usepackage{psfig}
\usepackage{epsf}
\usepackage{graphics}
\usepackage{lscape}
\def\edcomment#1{\iffalse\marginpar{\raggedright\sl#1\/}\else\relax\fi}
\reversemarginpar

\def\xmm{{{\it XMM-Newton}}}
\def\rosat{{{\it ROSAT}}}

\begin{document}
\title{XMM-Newton Observations of Extra-planar Gas in Nearby Starburst Galaxies}
\author{M. Ehle}
\affil{XMM-Newton Science Operations Centre, European Space Agency, Villafranca 
del Castillo, P.O. Box 50727, 28080 Madrid, Spain}

\begin{abstract}

We report on first results\footnote{
This work is based on observations obtained with \xmm, 
an ESA science mission with instruments and contributions directly 
funded by ESA Member States and the USA (NASA).
}
of \xmm\ observations of nearby starburst galaxies that form part of a 
multi-wavelength study of gaseous halos around late-type spiral galaxies and their dependence
on the level of star formation activity in the underlying disks.

\xmm, with its extraordinary sensitivity for faint extended X-ray emission, 
is used to derive spatial and spectral properties of the very hot extraplanar/halo 
gas. For example, spectral models can be tested and hot gas properties like
density, mass and energy can be estimated. Comparing the distribution of the halo
X-ray emission with optical filaments and/or observed magnetic field structures 
uncovers interesting correlations on which work just has started. 

Our study aims - in general - at assessing the importance of galactic halos as repositories 
of a metal-enriched hot medium and their significance in terms of galactic chemical 
evolution and possible metal enrichment of the intergalactic medium.

\end{abstract}
\thispagestyle{plain}

\section{Introduction}
We have selected nearby edge-on oriented starburst galaxies to perform a 
multi-wavelength (X-ray, radio-continuum, H{\sc{I}} and optical) study of all 
phases of extraplanar gas to investigate the dependence of galactic halo properties 
on the energy input rate due to star formation (SF) in the underlying galactic disk. 
Special care must be taken to separate halos created by a sufficiently high energy
input and extra-planar emission which is caused by galactic interactions or nuclear 
activity. The galaxy's size, i.e. the depth of the gravitational potential, is 
another parameter affecting the evolution of galactic halos. One of the main goals 
of this project is the determination of the energy budget in the halos (magnetic 
field, thermal and radiation energy densities).
                                                                                
\xmm\ (\citet{XMM}) X-ray observations presented here are used to detect previously
unknown extended halo emission due to hot gas and/or to re-visit galaxies with
known (from earlier X-ray missions) extra-planar X-ray emission features but now
with a hitherto unreached sensitivity that can only be provided by deep \xmm\ 
observations. Such observations also allow us to investigate with
unprecedented signal-to-noise the characteristics (temperature, metallicity, 
energy density) of the hot gas via X-ray spectroscopy.

An earlier report on the status of our ongoing work was given by
\citet{iaus217} and \citet{memorie}. A more detailed motivation and general 
introduction of our multi-wavelength project can be found in the paper by 
M. Dahlem (this volume).

\section{Observations}
This paper is based on \xmm\ observations of the starburst galaxies NGC~1511, 
NGC~4666 and NGC~3628 carried out as part of the Guaranteed Time proposal 011098
in July 2000, June 2002 and November 2000, respectively. The X-ray data have been
cleaned for periods of high radiation background and images and spectra were
generated with the Science Analysis System (SAS) software package (in its version 
5.4.1, except for NGC~1511 that was processed with 5.3.3). We also made use of the
EPIC background blank field files and tools provided by A. Read \citep{blankfield}.

\section{Results}

\subsection{NGC~1511}
The starburst SAa pec:H{\sc{II}} type galaxy NGC~1511 was studied as part of our 
project and X-ray results presented at this meeting have been published in 
\citet{n1511_xmm}.
\xmm\ EPIC (\citet{EPIC_pn,EPIC_MOS}) revealed for the first time the presence of 
a diffuse hot gaseous phase in NGC~1511 partly extending out of the disk plane. 
Extra-planar emission due to cosmic rays and magnetic fields was earlier seen in 
radio continuum emission \citep{rc_halos2} and is suggestive of a common origin 
for the outflow of these components of the ISM. The X-ray spectral
analysis of the integrated 0.2-12 keV emission (excluding a strong point
source about 30\arcsec\, north of the centre, which - if associated with
NGC~1511 - might be an ultra-luminous $L_X=1.18\,10^{40}$~erg s$^{-1}$ 
X-ray source) showed a complex emission composition: one (although not the only
possible) best-fitting model was found consisting of two thermal components and 
a power law contributing 12\% (0.19 keV), 11\% (0.59 keV) and 77\% (powerl) to 
the total flux, respectively. The finding that a spectral model with a single 
temperature gas component is not sufficient to describe the emission, points 
toward the fact that the X-ray emitting gas contains several phases. The best-fit 
model corresponding total X-ray luminosity ($L_X=1.11\,10^{40}$~erg s$^{-1}$) leads 
to a far-infrared-to-X-ray luminosity ratio for NGC~1511 which is typical for 
starburst galaxies (\citet{heckman90,read+ponmanII}).

The Optical Monitor (OM, \citet{OM}) on board \xmm\ was used to observe NGC~1511 in 
the UV during the X-ray observations and obtained images showing that this galaxy 
is heavily disturbed (as it also can be seen in H$\alpha$ (\citet{lehnert+heckman95}) 
and in the near infra-red). Strong evidence for tidal interactions has only recently 
been obtained through H{\sc{I}} observations \citep{hi_halos}, unfortunately 
rendering NGC~1511 unsuitable as target galaxy to study the dependence of its 
gaseous halo's properties on the distribution and level of SF in the underlying 
disk.

\subsection{NGC~4666}
Based on multi-wavelength (optical, radio continuum and \rosat\ X-ray observations), 
we `classified' NGC~4666 as a `superwind' Sc galaxy \citep{n4666_multi}
harboring an extra-planar outflow cone emanating from a central SF region of 
$\sim3.2$~kpc in radius, and having an opening angle of $30\deg\pm10\deg$. The
outflow could be traced up to $\sim7.5$~kpc above the disk plane by optical emission
line filaments. 

\begin{figure}[!ht]
\plotone{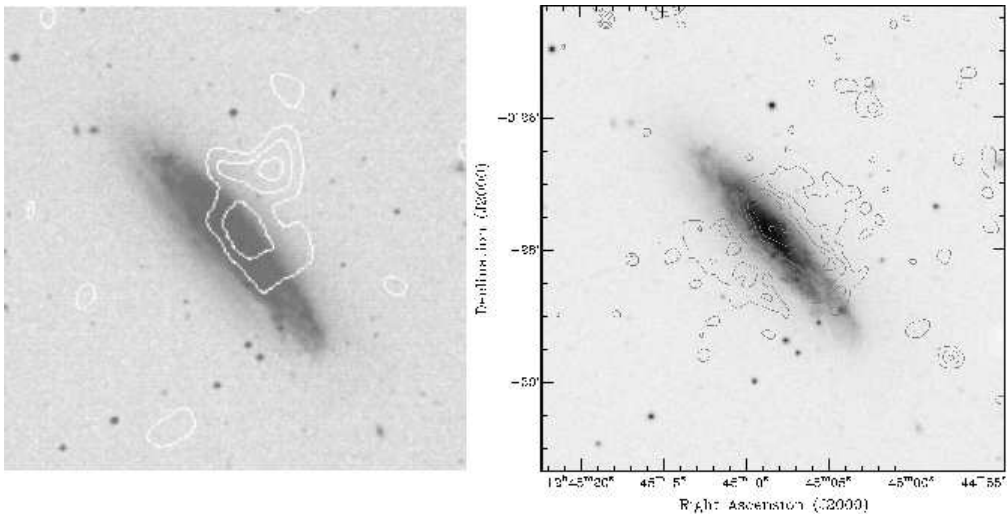}
\caption{Soft X-ray image of NGC~4666 obtained with \rosat\ PSPC at 0.25 keV ({\it
left panel}; from \citet{n4666_rosat}) and with \xmm\ EPIC ({\it right panel}) from 
combined pn-MOS data in the 0.2 - 0.5~keV energy band, overlaid on a DSS image. 
The \xmm\ data was slightly smoothed with a non-adaptive Gaussian to a spatial 
resolution of 10\farcs4.}
\label{fig_n4666_xray}
\end{figure}

\begin{figure}[!ht]
\plotfiddle{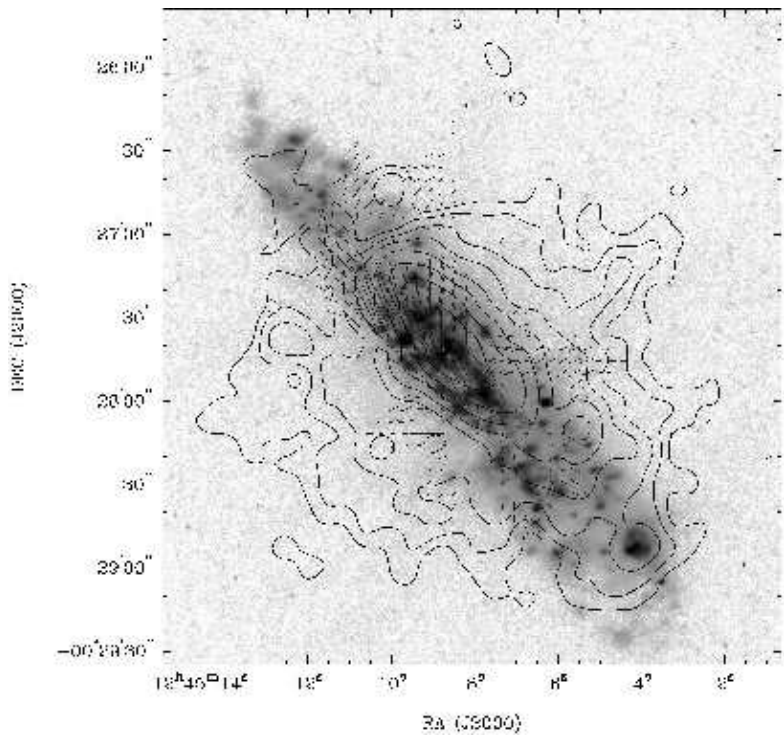}{9cm}{0}{50}{50}{-170}{-10}
\caption{\xmm\ combined pn-MOS image of NGC~4666 in the 0.5 - 0.9~keV 
energy band overlaid on the H$\alpha$+N$[\rm{II}]$ optical narrow band image 
from \citet{n4666_ha}. In the halo the most extended optical emission line
filaments form an ``X''-shaped structure (cf. the plates in \citet{n4666_hadeep}). 
Vectors mark the orientation of the magnetic field observed at 4.89 GHz with the 
VLA, their lengths are proportional to the polarized intensity.}
\label{fig_n4666_multi}
\end{figure}

The \rosat\ X-ray observations (\citet{n4666_rosat}, see Fig.~\ref{fig_n4666_xray}, 
left panel) indicated the presence of soft
extended X-ray emission outside of the disk on the north-western (closer to us) side
of NGC~4666. Our \xmm\ EPIC observation, in contrast, show for the first time
that such soft X-ray emission exists on both sides of the galactic disk, originating
from a huge, structured hot gas halo (Fig.~\ref{fig_n4666_xray}, right panel, and 
Fig.~\ref{fig_n4666_multi}).

\begin{figure}[!ht]
\plotfiddle{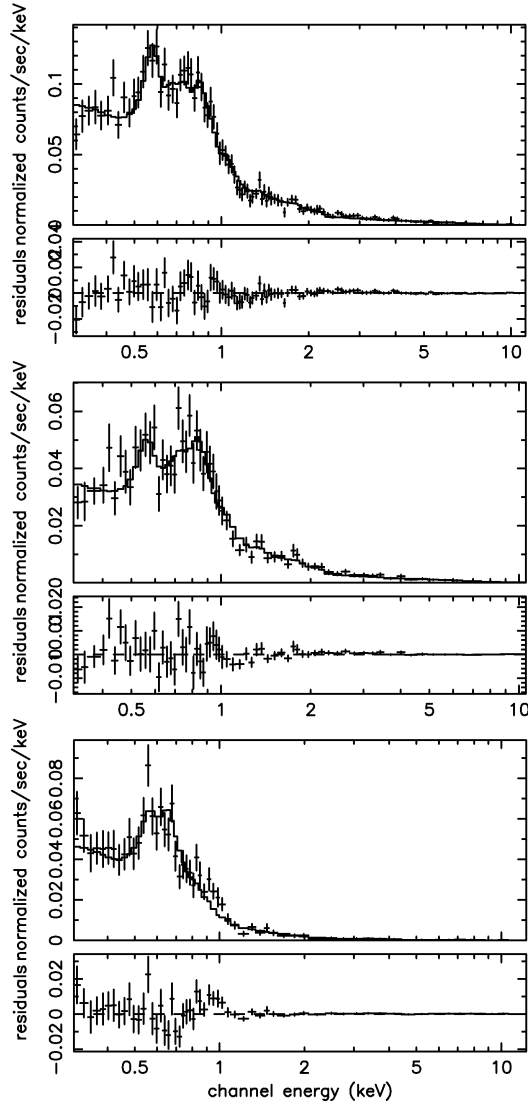}{14cm}{-90}{30}{30}{-115}{420}
\caption{XMM-Newton EPIC-pn spectrum of the disk ({\it top}), lower halo ({\it 
middle}) and upper halo ({\it bottom}) region of NGC~4666. The spectral models 
are described in the text. In the lower panels, residuals between the spectral 
model and the data are shown.}
\label{fig_n4666_pnspectrum}
\end{figure}

The diffuse emission detected by EPIC-pn is strong enough to allow us to perform 
a spectral analysis in several areas in the galactic disk and halo: the complex 
spectrum of the diffuse disk emission (Fig.~\ref{fig_n4666_pnspectrum}, top) 
can be fitted by a combination of an internally absorbed MEKAL (0.54~keV) and power 
law component plus another MEKAL model (0.18~keV) all affected by Galactic 
foreground absorption. The lower halo emission (Fig.~\ref{fig_n4666_pnspectrum}, 
middle) shows a similar spectral behavior as the disk spectrum. The spectrum of the
upper halo emission (Fig.~\ref{fig_n4666_pnspectrum}, bottom), however, does not 
need a second MEKAL component but is fitted reasonably well also with a single 
0.23~keV thermal plasma. Most of the total flux above 0.9~keV originates from the 
power-law type emission (presumably due to unresolved point-like sources), whereas 
the thermal plasma clearly dominates in the soft 0.3-0.9~keV band 
(contributing 72\%, 69\% and 92\% to the `diffuse' disk, lower and upper halo 
emission, respectively). 

If the diffuse X-ray emission in the soft energy band is assumed to be
due to hot gas, it is possible to calculate the gas density $n_{\rm e}$
and its mass $m_{\rm gas}$. To this end we used the model of thermal cooling 
and ionization equilibrium of \citet{nulsen84} where 
$L_{\rm x}({\rm soft})\simeq\Lambda(T) n_{\rm e}^2 V \eta$. The unknown 
filling factor $\eta$ allows for some clumpiness of the gas. 
If we fit all the spectra extracted from the three areas with a 
single-temperature MEKAL model, the unabsorbed flux in the 0.3 - 12~keV and 
corresponding luminosity (adopting $D=26.3$~Mpc) are given in Tab.~\ref{tab_n4666}. 
For the gas temperatures of the hot gas, \citet{rasm76} give a cooling 
coefficient $\Lambda(T)$ of $\sim10^{-22}$ erg cm$^3$ s$^{-1}$.
Assuming an ellipsoid for the disk and the outflow cone geometry (see above) for the 
emitting halo volume $V$, and typical volume filling factors of 0.1 - 0.8, we derive 
gas densities and masses listed in Tab.~\ref{tab_n4666}. 

\begin{table}[!ht]
\caption{Parameters of the diffuse X-ray emission components of NGC~4666}
\label{tab_n4666}
\smallskip
\begin{center}
{\scriptsize
\begin{tabular}{lcccccc}
\tableline
\noalign{\smallskip}
           & T     & $f_{X,0.3-12}^{mekal,unabs}$ & $L_X$ & $n_e$ & $m_{Gas}$ & $U_{th}/U_{mag}$\\
           & {\tiny [keV]} & {\tiny [10$^{-14}$~erg/cm$^2$/s]} & {\tiny [10$^{39}$~erg/s]} & {\tiny [10$^{-3}$cm$^{-3}$]} & {\tiny [10$^7 M_{sun}$]} & \\
\noalign{\smallskip}
\tableline
\noalign{\smallskip}
disk       & $0.29\pm0.01$ & 7.21 & 5.97 &  1.0 - 2.9 & 2.5 - 7 & 0.1 - 0.2 \\
lower halo & $0.35\pm0.03$ & 2.77 & 2.29 &  2.5 - 7.1 & 0.5 - 1 & 0.7 - 2.0 \\
upper halo & $0.23\pm0.01$ & 4.40 & 3.64 &  1.2 - 3.3 &   1 - 4 & 0.2 - 0.6 \\
total halo & $0.27\pm0.01$ & 7.38 & 6.11 &  1.4 - 4.0 &   2 - 5 & 0.3 - 0.9 \\
\noalign{\smallskip}
\tableline
\end{tabular}
}
\end{center}
\end{table}

Such estimated gas densities together with the fitted temperature allow us to 
compare the energy density of the hot gas with e.g. that of the magnetic field
(magnetic field strengths were derived from radio continuum observations 
as 14.4~$\mu$G in the disk and 7.1~$\mu$G in the halo (see \citet{n4666_multi}).
Whereas in general the ratio between the thermal and magnetic energy densities
is $<1$ (see Tab.~\ref{tab_n4666}), and hence the magnetic field important
for the dynamics of the hot gas (`channeling' the outflow), in the lower halo
the relatively high thermal energy density is supportive of the idea that
in this region the gas outflow might easily take place, even against a
disk-parallel magnetic field configuration.

A detailed analysis of the halo emission and a possible correlation with
H$\alpha$ and radio polarization filaments (as started in \citet{n4666_multi},
see also Fig.~\ref{fig_n4666_multi}) will be addressed in an upcoming paper by 
Ehle et al.; there we also plan to present a self-consistent dynamical and thermal
spectral model \citep{breit99,breit03} which does no longer depend on the usual 
assumption of collisional ionization equilibrium (CIE). 

We note that the same \xmm\ observations have been discussed by 
\citet{n4666_XMM+SAX} comparing 
the disk-emission with spatially unresolving {\it BeppoSAX} data. The authors 
discuss their findings with respect to the discovery of starburst plus AGN 
activity both contributing to the X-ray emission of NGC~4666: 
the SF activity was found to be extended over most of the disk and associated with 
diffuse thermal emission, whereas the quite small low-luminosity type-2 AGN 
contribution (revealed by prominent K$\alpha$ line emission from ``cold'' iron 
at 6.40~keV) was detected from the nuclear region.
   
In NGC~4666, \citet{n4666_hi} failed to detect conclusive evidence for the 
existence of H{\sc{I}} gas in its halo. However, seen in their H{\sc{I}} maps 
are prominent tidal arms probably generated by interactions with the galaxy 
NGC~4668 and previously undetected dwarf companions. The high SF rate responsible 
for the superwind of NGC~4666 hence might be triggered by gravitational 
interaction.  

\subsection{NGC~3628}
\begin{figure}[!ht]
\plottwo{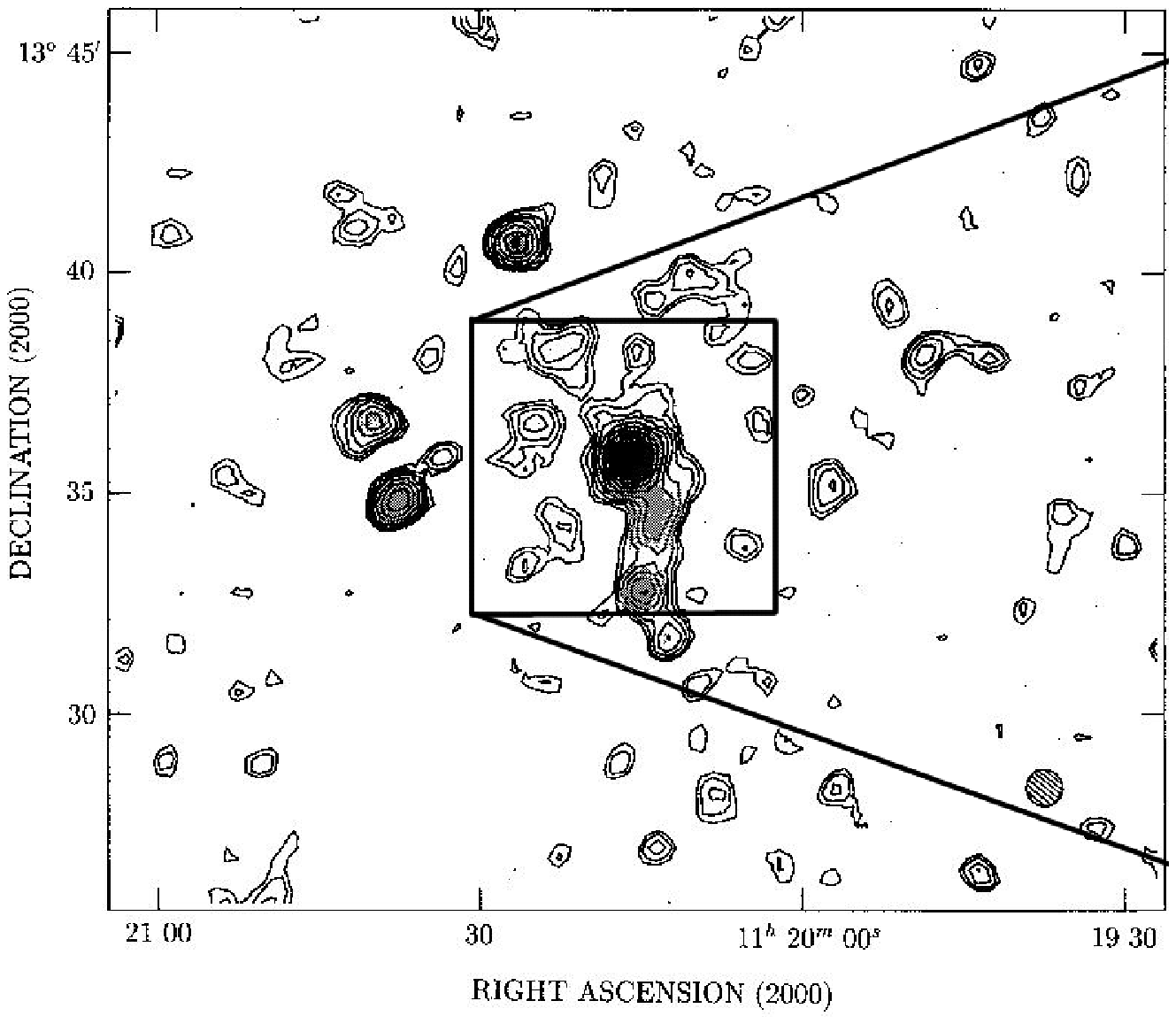}{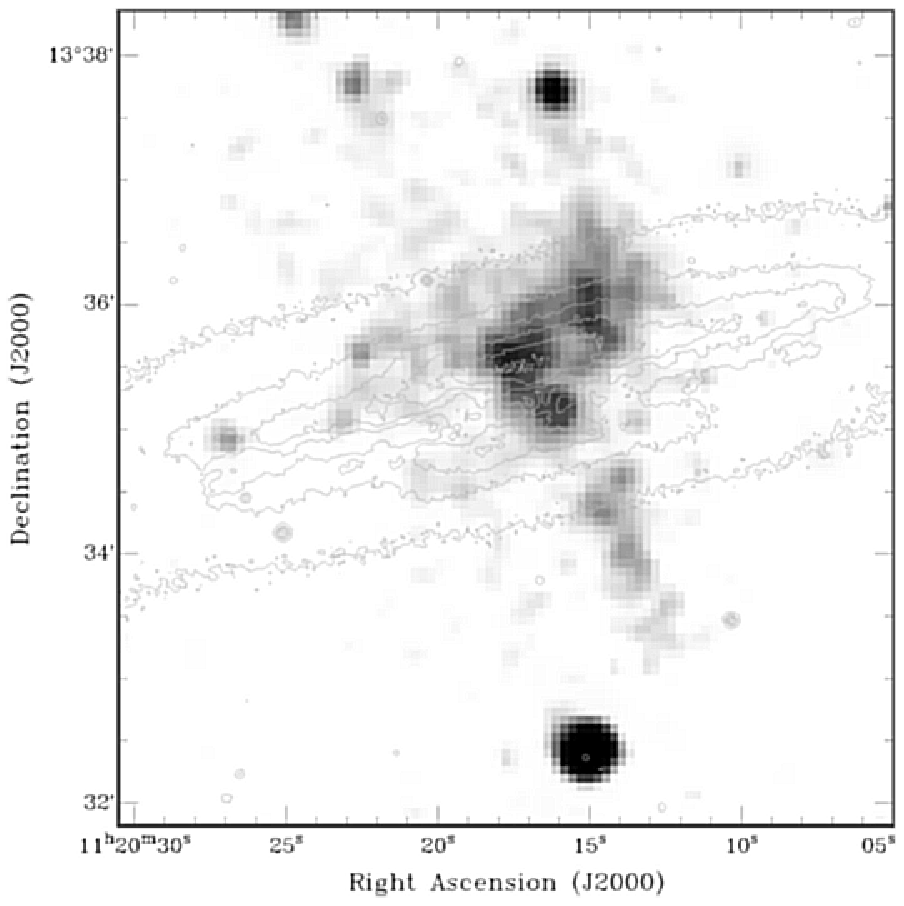}
\caption{
X-ray image of NGC~3628 obtained with \rosat\ PSPC at 0.75 keV ({\it
left panel}; from \citet{n3628_rosat}) and with \xmm\ EPIC ({\it right panel}) 
from combined pn-MOS data in the 0.3 - 2.0~keV energy band shown as greyscale 
image with overlaid contours of a DSS image. 
Whereas the \rosat\ image has a spatial resolution of 48\arcsec, the \xmm\ data 
was slightly smoothed with a non-adaptive Gaussian to 10\arcsec. The `box' 
painted on top of the \rosat\ map roughly marks the area of the displayed EPIC 
image. 
}
\label{fig_n3628_rosat+xmm}
\end{figure}

NGC~3628 is a peculiar Sbc galaxy known to be an interacting member in the 
Leo Triplet (Arp 317, assumed distance $D=10$~Mpc). Earlier X-ray observations 
with {\it Einstein} \citep{n3628_einstein} and \rosat\ (\citet{n3628_rosat}, 
see Fig.~\ref{fig_n3628_rosat+xmm}, left panel) 
showed evidence for a collimated outflow along the minor axis from a starburst 
nucleus in NGC~3628 and led to the detection of an extended soft X-ray halo.

\begin{figure}[!ht]
\plottwo{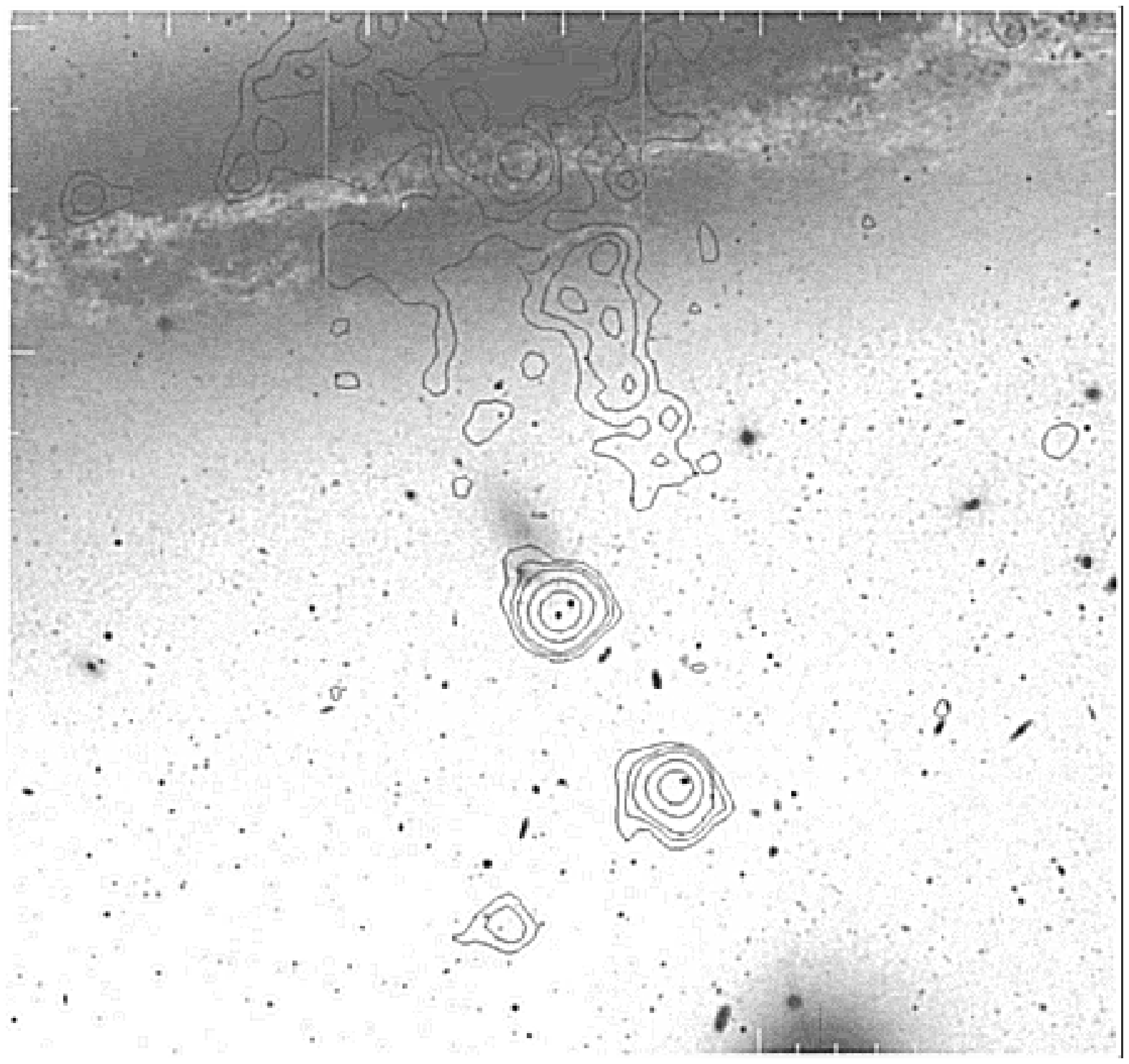}{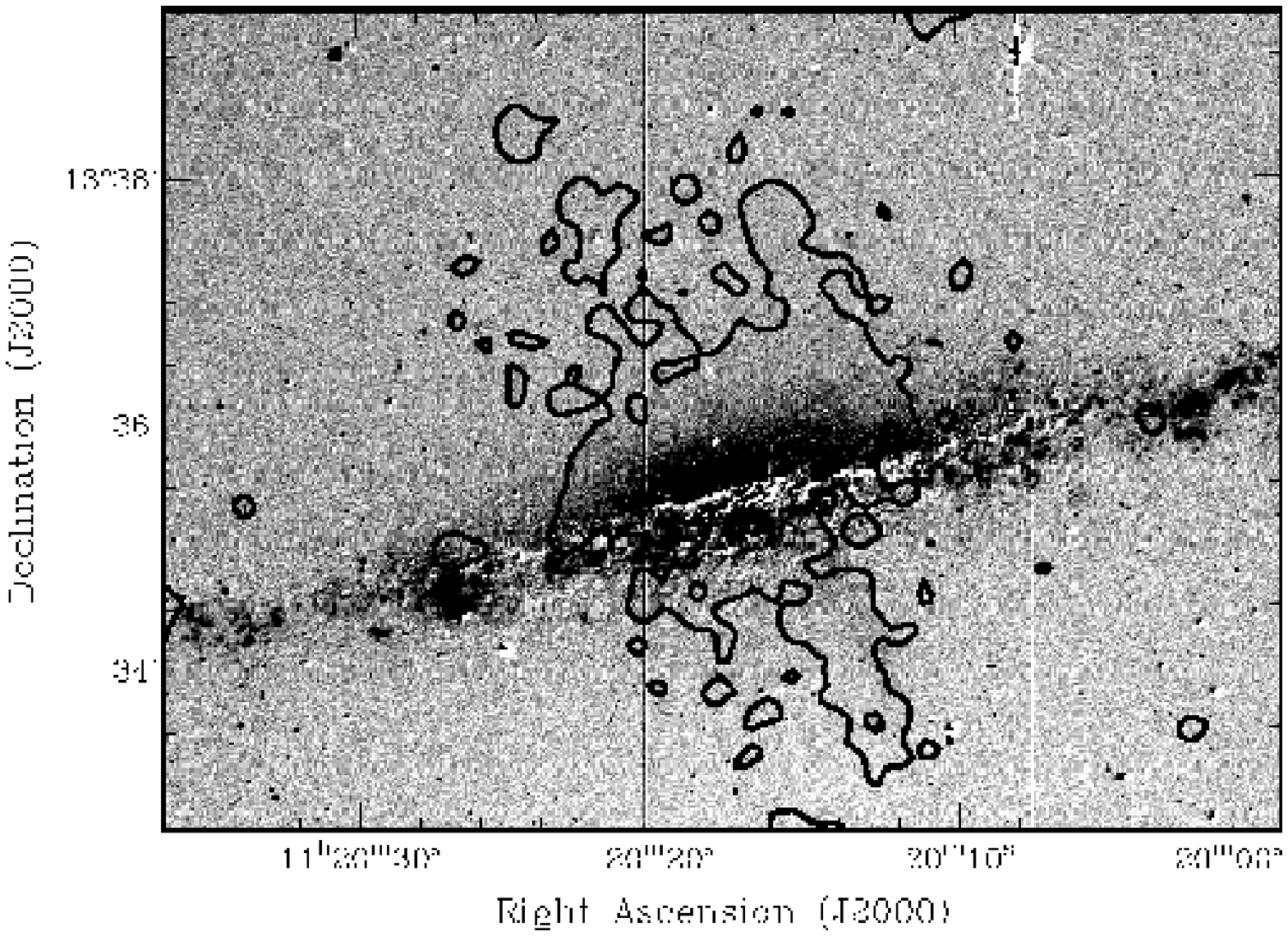}
\caption{
\xmm\ EPIC 0.5-0.9~keV contour map of NGC~3628 over plotted on a VLT-FORS2 image 
({\it left}, see ESO Messenger, March 2002) and (only showing a single low 
intensity contour) on the H$\alpha$ greyscale image (from \citet{n3628_halpha}).
The `box' in the left panel roughly marks the spatial extent of the {\it Chandra}
image by \citet{n3628_chandra}, their Fig.~1).}
\label{fig_n3628_xmm+opt}
\end{figure}

Our XMM-Newton observations (Fig.~\ref{fig_n3628_rosat+xmm}, right panel) are able 
to detect the extraplanar diffuse emission with much higher significance: 
the EPIC image clearly separates the southern collimated spur-like halo 
emission from nearby (most possibly background) point sources (see 
Fig.~\ref{fig_n3628_xmm+opt}, left panel) and calls the 
proposed link between this X-ray filament and QSOs \citep{arp02} to question.

A detailed comparison of the diffuse extraplanar X-ray emission with for example
the H$\alpha$ filaments (a plume extending about 130\arcsec\, to the SW in 
position angle $\sim210\deg$ and faint more widespread filamentary 
extraplanar structures to the north - see H$\alpha$ map from 
\citet{n3628_einstein} or \citet{n3628_halpha} and 
Fig.~\ref{fig_n3628_xmm+opt}, right panel) is ongoing. 

\section{Discussion}
In the framework of our multi-wavelength project to investigate gaseous halos around
late-type spiral galaxies and their dependence on the level of star-formation activity
in the underlying disks (cf. M. Dahlem (this volume)) we showed in this paper
X-ray results from three of our sample galaxies. The presented observations 
demonstrate that only with the advent of high sensitivity observatories (like \xmm), 
it is becoming possible to detect and study in detail the hot halo gas component 
of the interstellar medium. 

Our original sample selection criterion (based on far infra-red colors, see
\citet{rc_halos2} and references therein), that showed to be a good tool
to select candidates for the search for radio halos, is likely to also work
in the X-ray regime. In all three \xmm\ targets we find extra-planar X-ray emission
which is bright enough to allow us detailed spectral investigations and also
to test model assumptions (i.e. collisional ionization equilibrium versus radiative
cooling with the dynamics in full non-equilibrium).

We note the co-existence of extra-planar soft X-ray emission above the most
actively star forming regions in the galactic disk and the co-existence
of such features with H$\alpha$ filaments as well as vertical magnetic field
structures (in case of NGC~4666).

\end{document}